\documentclass[%
 reprint,
 amsmath,amssymb,
 aps,
pre,
]{revtex4-2}

\usepackage{graphicx}
\usepackage{dcolumn}
\usepackage{bm}

\begin{document}

\preprint{APS/123-QED}

\title{Elastocapillary adhesion of soft gel microspheres}

\author{Joseph N. Headley}  
\altaffiliation[These authors contributed equally to this work.]{ }
\affiliation{Department of Physics and Astronomy, Williams College, Williamstown, MA 01267}
\author{Edgar W. Lyons} 
\altaffiliation[These authors contributed equally to this work.]{ }
\affiliation{Department of Physics and Astronomy, Williams College, Williamstown, MA 01267}
\author{Mathew Q. Giso} 
\altaffiliation[These authors contributed equally to this work.]{ }
\affiliation{Department of Physics and Astronomy, Tufts University, Medford, MA 02155}
\author{Emily P. Kuwaye} 
\affiliation{Department of Physics and Astronomy, Williams College, Williamstown, MA 01267}
\author{Caroline D. Tally} 
\affiliation{Department of Physics and Astronomy, Williams College, Williamstown, MA 01267}
\author{Aidan J. Duncan} 
\affiliation{Department of Physics and Astronomy, Williams College, Williamstown, MA 01267}
\author{Chaitanya Joshi} 
\affiliation{Department of Physics and Astronomy, Tufts University, Medford, MA 02155}
\author{Timothy J. Atherton}
\email{timothy.atherton@tufts.edu}
\affiliation{Department of Physics and Astronomy, Tufts University, Medford, MA 02155}
\author{Katharine E. Jensen}
\email{kej2@williams.edu}
\affiliation{Department of Physics and Astronomy, Williams College, Williamstown, MA 01267}

\date{\today}

\begin{abstract}
Softer means stickier for solid adhesives, because material compliance facilitates close contact between non-conformal surfaces.
Recent discoveries have revealed that soft materials can exhibit a rich array of new physics arising from competing effects of continuum elasticity, fluid-like surface mechanics, and internal poroelastic flows, all of which can directly impact interfacial interactions.
In this work, we investigate this complex interplay across several orders of magnitude of elastic stiffness by measuring the complete adhesive contact geometry between compliant silicone gel microspheres and flat, rigid substrates.
We observe a continuous elastocapillary transition in adhesion mechanics, with novel features revealed by both the breadth of data and the detailed contact geometries.
Importantly, soft gel spheres exhibit a remarkably broad range of near-equilibrium contact morphologies and their contact line deformation is always mediated by a fluid contact zone that phase separates from the gel.
To explain this, we develop a new model incorporating elastocapillary and poroelastic mechanics that predicts the complete range of adhesive behavior and elucidates energetic tradeoffs. The data and model together reveal a shallow energy landscape that may contribute to the robustness of everyday adhesives. 
\end{abstract}

\maketitle


\section*{Introduction}

We depend on the soft and sticky: from everyday applications like adhesive tapes, bandages, and non-slip surfaces to large-scale manufacturing, construction and cutting-edge innovations in soft robotics, flexible electronics, sutures, and biomedical devices \cite{creton2003pressure,he2023adhesive,Persson2001,Kim2007,shintake2018soft,gong2017toward}, contacts between compliant and uneven surfaces are essential to everyday life, medicine, and technologies.
Two surfaces tend to adhere to each other if they have a positive adhesion energy per area, $W = \gamma + \gamma_1 - \gamma_{\text{int}}$, where $\gamma_{\text{int}}$ is the interfacial free energy and $\gamma$ and $\gamma_1$ are the surface energies per area of each individual free surface, respectively. However, establishing a strong contact between initially non-conformal surfaces can require significant deformation at the interface and hence the mechanical properties of interacting solids play a critical role in determining how effectively an adhesive contact can be made.

Softer materials tend to be stickier because they can more easily deform into conformal contact \cite{ca1966tack, creton2003pressure,creton2003materials,creton2016fracture,jensen2025physics}.
Indeed, a key design principle for pressure-sensitive adhesives (PSAs), the Dahlquist Criterion, states that an effective PSA should have a shear modulus $<100$ kPa \cite{ca1966tack}. 
However, soft solids are often highly complex, and can exhibit properties typical of both solids and liquids. 
Classical theories do not always account for this complexity: for example, the classic Johnson, Kendall, and Roberts (JKR) theory of contact mechanics \cite{JKR1971,Maugis1995,maugis2000contact} considers only the competition of adhesion energy and bulk elasticity in determining adhesive deformation. 
Significant work over the past 15 years has shown that solid surface tension can emerge as the dominant restoring stress in very soft solids on small length scales \cite{Mora2010,mora2011surface,Jerison2011,jagota2012surface,style2013surface,style2017elastocapillarity,bico2018elastocapillarity,andreotti2020statics,heyden2024distance,tamim2025shaping}.

Compliant solids in this \emph{elastocapillary} regime demonstrate contact behavior more typically associated with liquids, including particle adhesion that quantitatively resembles adsorption on a liquid surface \cite{style2013surface,cao2014adhesion} and anomalously high adhesion due to solid capillary forces \cite{jensen2017strain,liu2019effects}.
Many highly compliant materials are gels: microscopically heterogeneous materials composed internally of a percolated crosslinked elastic network permeated by a free fluid phase. Mounting evidence shows that poroelastic \cite{hu2012viscoelasticity} flow of the gel fluid phase plays an essential role in governing gel mechanics, from phase-separated static mechanical equilibria \cite{jensen2015wetting, liu2016osmocapillary, glover2020capillary} to highly dynamic processes \cite{karpitschka2015droplets, berman2019singular, xu2020viscoelastic,jeon2023moving,hauer2023phase}. 
Together, these results suggest that the contact mechanics of compliant gel surfaces are determined by a complex interplay of surface, bulk, and internal fluid mechanics that remains challenging to access experimentally.

In this work, we image the contact geometry of soft gel microspheres in adhesive contact with flat, rigid substrates across six orders of magnitude of gel stiffness. This geometry is of considerable importance since many practical adhesives---including the ubiquitous sticky note---are textured with soft, sticky asperities on the scale of 10s to 100s of micrometers \cite{creton2003pressure,jensen2025physics}. 
Only a few prior studies have examined this geometry \cite{joanny2001gels,carrillo2010adhesion,Salez2013,chakrabarti2018elastowetting}, in contrast to numerous studies of the reciprocal case with rigid spheres and compliant substrates \cite{style2013surface, Xu2014, cao2014adhesion, jensen2015wetting, hui2015indentation, jensen2017strain, xujensen2017direct, pham2017elasticity, ina2017adhesion}. 
While both geometries possess convergent mechanics in the small deformation limit, compliant spheres are able to access much larger deformations. 

We observe a transition from classical elastic- to capillary-dominated adhesion with decreasing microsphere stiffness and size. In all adhesion regimes, we find that gel fluid phase separation plays an essential role in mediating adhesive contact. 
Expanding on earlier theoretical work \cite{Salez2013}, we develop a minimal model that incorporates adhesion, elasticity, capillarity, and gel fluid phase separation and converges on the predictions of classical models, such as JKR, in appropriate limiting cases. 
The model accurately captures both the overall deformation behavior and the observed adhesion morphologies, and enables us to untease the system's tradeoffs  between competing terms in the total energy. 
The existence of shallow directions in the energy landscape about equilibrium for an adhesive contact may facilitate more versatile contacts and explain the robustness of practical adhesives. 
By characterizing the adhesive behavior of compliant gel asperities across such a large parameter space, our results expand our fundamental understanding of adhesion with soft gels, and emphasize the critical role of both solid and liquid capillarity in soft adhesion.

\begin{figure*}[t!]
    \centering
    \includegraphics{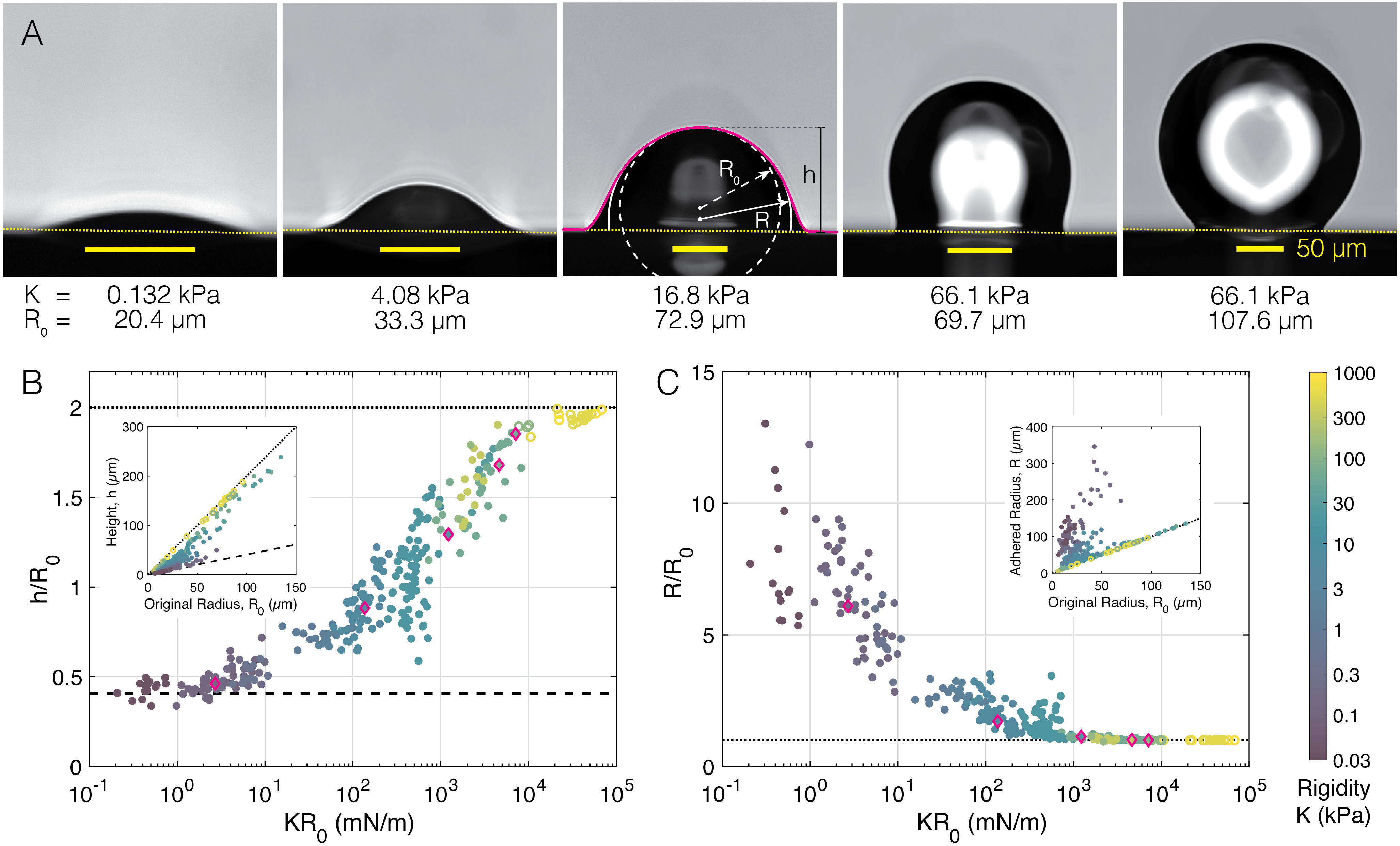}
    \caption{\textbf{Adhesive contact geometry.}  
    \textit{(A) } Example raw data images across a range of rigidity $K$ and initial radii $R_0$. Yellow dotted line indicates flat, rigid surface to which the microspheres adhere. 
    \textit{Center overlay:} 
    Mapping the edge profile (magenta) of the flat surface and adhered sphere yields the complete microsphere geometry, including height $h$, adhered radius $R$ (white line), and original sphere radius $R_0$ (white dashed line).
    We approximate the foot volume as the region between the extrapolated central spherical cap with radius $R$ and the measured profile.
    \textit{(B) } Extent of deformation $h/R_0$ vs. $K R_0$ for 293 individual microspheres adhered to bare Pyrex glass substrates. 
    \textit{Inset:} Adhered height $h$ vs. original radius $R_0$.
    \textit{(C) } Normalized adhered radius $R/R_0$ vs. $K R_0$.
    \textit{Inset:} Adhered radius $R$ vs. $R_0$.
    In \textit{(B-C)}, dotted lines indicate the geometric bound of undeformed spheres, while dashed lines show a fit to the empirically-observed bound of maximum deformation at $h/R_0 = 0.41$.
    Open symbols indicate data for which $R\approx R_0$; see Materials and Methods.
    Magenta diamonds highlight the points corresponding to the examples shown in Ref. \textit{(A)}.
    Colormap from \cite{biguri2024perceptually}.
    }
    \label{fig:contact_geometry}
\end{figure*}

\section*{Adhesive deformation of gel microspheres}

To characterize their adhesive behavior experimentally, we prepare a library of polydimethylsiloxane (PDMS) gel microspheres over a broad range of size and stiffness. 
We fabricate silicone gel microspheres by mechanically emulsifying a mixture of reactive PDMS linear polymer, crosslinker, and Platinum Karsted's catalyst together in deionized (DI) water, and then allow the emulsified droplets to crosslink covalently into solid gel spheres at room temperature for at least 24 hours.
In the absence of crosslinking, the PDMS linear polymer liquid totally wets bare glass ($\gamma_l = 20$ mN/m, $W \geq 2\gamma_l$), and consequently the gels are very sticky.

The microspheres thus produced are highly polydisperse, ranging from several $\mu$m to around 150 $\mu$m in diameter. 
By adjusting the polymer to crosslinker ratio, we fabricate solid gels with varying rigidity $K = \frac{4 E}{3 (1-\nu^2)}$, where $E$ is Young's modulus and $\nu \approx 0.48$ \cite{jensen2015wetting} is Poisson's ratio \cite{Maugis1995,Salez2013}.
We conduct experiments with gels spanning from $K = 0.032$ kPa to $K = 870$ kPa, ranging from a rigidity comparable to the softest biological tissues to that of a stale gummy bear. 
The fluid phase of the gel comprises uncrosslinked polymer that remains after the crosslinking reaction is complete, and all samples have a substantial free fluid fraction that decreases with increasing stiffness (see Supporting Information Table S1 and Fig. S2).
More details on gel fabrication and characterization are included in the Materials and Methods section.

We allow the microspheres to establish adhesive contact with flat, rigid glass substrates, and then image hundreds of individual, adhered spheres using brightfield microscopy as described in Materials and Methods.
Example raw data images are shown in Fig. \ref{fig:contact_geometry}\textit{(A)}. 
In all cases, the adhered geometry takes the form of a central spherical cap with a meniscus or ``foot'' \cite{joanny2001gels,chakrabarti2018elastowetting} that bridges between the spherical cap and the flat glass substrate; 
the feet comprise the largest fraction of the total volume in the intermediate-compliance regime. 
In contrast to previous studies that considered a foot resulting from substantial elastic deformation of the solid gel \cite{chakrabarti2018elastowetting}, here the foot is a phase-separated fluid as we determine with confocal microscopy (see Supporting Information Fig. S3).

From such raw images, we measure the profile of each microsphere and the position of the flat glass plane with a resolution of 100 nm using custom edge-mapping Matlab software developed in our earlier works \cite{jensen2015wetting,jensen2017strain,xujensen2017direct}, as indicated in Fig. \ref{fig:contact_geometry}\textit{(A, center overlay}). 
Because the spheres adhere axisymmetrically, confirmed using both top-view brightfield and confocal microscopic imaging, these side-view profiles allow us to extract the complete contact geometry of each microsphere (Fig. \ref{fig:contact_geometry}\textit{(A, center overlay})). 
This includes the adhered radius $R$ of the central spherical cap, the adhered height above the flat substrate $h$, and the original sphere radius $R_0$ prior to adhesion (inferred using conservation of volume for the entire axisymmetric shape).
We also measure both the surface angle relative to the horizontal plane of the substrate and the total curvature $C$ \cite{deGennes2004} along each sphere profile by fitting locally with a surface of constant total curvature, as developed in our previous work \cite{jensen2015wetting}. 
For every profile, the curvature is symmetric about the center of axisymmetry (see Supporting Information Figs. S4 and S5), with $C = 2/R$ in the central region region, as expected for a spherical cap, and with sharp regions of approximately constant negative total curvature in the feet, indicative of a negative (tensile) Laplace pressure \cite{deGennes2004} in the phase-separated fluid.

By measuring contact geometries over such a large parameter space, we gain detailed insight into how soft gel asperities of different size and stiffness accommodate an adhesive interaction with a rigid substrate.
At one extreme, the largest and stiffest spheres adhere with only very small deformations, such that the adhered radius $R \approx R_0$, while the smallest and most compliant microspheres are deformed substantially in adhesive contact, with $R \gg R_0$. 
We plot $h$ vs. $R_0$ and $R$ vs. $R_0$ for 293 individual adhered microspheres as insets in Fig. \ref{fig:contact_geometry}\textit{(B)} and \textit{(C)}, respectively. 
Since both height and adhered radius generally scale with initial sphere size, we quantify the extent of deformation of these spheres by plotting $h/R_0$ in Fig. \ref{fig:contact_geometry}\textit{(B)} and $R/R_0$ in Fig. \ref{fig:contact_geometry}\textit{(C)}. 
In both cases, we show the extent of deformation versus initial radius scaled by the rigidity, $K R_0$.
This rescaling collapses all of the data into a broad curve bounded at the high $K R_0$ extreme by the geometry of an undeformed sphere with $h/R_0 = 2$ and $R/R_0 = 1$. 
We observe a large variation in extent of deformation even for spheres of the same size and stiffness that significantly exceeds what could be accounted for by potential small sphere-to-sphere variations in material properties, especially in the intermediate-compliance regime ($KR_0 \sim 10^2-10^3$ mN/m).

The deformation data also reveal a plateau at low $K R_0$ of constant $h/R_0 = 0.41$ for the smallest and most compliant spheres, measured by fitting the smallest decade of $K R_0$ data, despite the fact that the uncrosslinked fluid is totally wetting on the glass. 
From a wetting perspective, each $h/R_0$ value also corresponds uniquely to a constant macroscopic contact angle $\Theta$, defined as the interior angle where the extrapolated central spherical cap would intersect the flat substrate, in direct analogy to partially-wetting sessile fluid droplets.
In the case of fluid droplets, the equilibrium contact angle is determined by the competition between adhesion (or wetting) and surface tension energies via the Young-Dupr\'e equation, $\cos\Theta=W/\gamma-1$, and connected to $h/R_0$ by volume conservation. 
From this perspective, the plateau at $h/R_0 = 0.41$ at the limit of vanishing elasticity in our data can be thought of as a capillary limit with $\Theta = 18^\circ$ and $W/\gamma = 1.95$.
Although the macroscopic contact angle grows significantly from the softest to stiffest spheres, with $18^\circ \leq \Theta \leq 180^\circ$, we find that the microscopic contact angle $\theta$ where the edge of the foot meets the flat substrate is constant and approximately zero ($\theta = 6^\circ \pm 9^\circ$) for all measurements, consistent with our observation that the foot comprises phase-separated fluid that totally wets the glass (see Supporting Information Fig. S6).

\section*{Extended elastocapillary model}

\begin{figure}[t!]
    \centering
    \includegraphics[width=1\columnwidth]{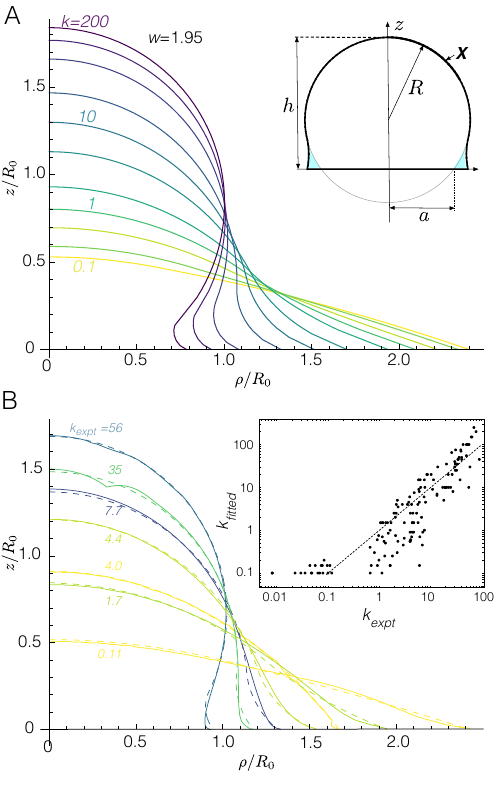}
    \caption{\textbf{Extended elastocapillary model.}  
    \textit{(A)} Extended elastocapillary model predictions for equilibrium adhesion geometries as a function of normalized stiffness $k\in[0.1,0.2,0.5,1,...,200]$ for adhesion $w=1.95$. \emph{Inset:} Geometry for the extended elastocapillary model, incorporating a compliant gel spherical cap with fluid feet. 
    \textit{(B)} Matched experimental (solid lines) and model (dashed) lines from fitting process. \emph{Inset:} Comparison of fitted and experimental stiffness $k$ from matched solutions, with $k_\text{fitted} = k_\text{expt}$ line indicated (black, dashed).
    }
    \label{fig:model}
\end{figure}

The overall transition of our data from a rigidity-dependent extent of deformation over high to intermediate $K R_0$ values to a quantitatively wetting-like plateau at low $K R_0$ can be captured by elastocapillary adhesion theory \cite{style2017elastocapillarity,style2013surface,Salez2013} that energetically bridges between classic elastic JKR adhesion \cite{JKR1971,maugis2000contact} and Young-Dupr\'{e} wetting \cite{deGennes2004}.
Particularly relevant is the Tenso-Elastic-Adhesive (TEA) model \cite{Salez2013} that sought to describe adhesion of soft elastic spheres to flat, rigid substrates assuming a perfect spherical cap adhesion geometry (see Supporting Information Text).
However, while such an elastocapillary model can capture the general trend of the deformation data, it fails to describe the experimentally-observed adhesion geometry of individual asperities and does not yet address the large variation in the extent of deformation measured even for nominally very similar microspheres.

In order to understand these experimental results, we develop a new model that expands this existing elastocapillary adhesion theory \cite{style2013surface,Salez2013} to incorporate additional essential physical mechanisms that are broadly relevant to many soft material systems. 
Our extended elastocapillary (XEC) model incorporates elasticity, adhesion, and surface tension with gel fluid phase separation such that a fraction of the internal fluid phase is able to separate out to form the foot contact zone. 
Qualitatively, this allows an adhering microsphere to achieve a larger area of contact and to have a vanishing contact angle at the edges of the foot without requiring an extreme elastic deformation of the solid gel. 
The liquid meniscus bridges between the substrate and the remaining, central solid gel.

Our extended elastocapillary model, shown schematically in Fig. \ref{fig:model}\textit{A (inset)}, is physically motivated by the experimentally-observed geometry.
Consistent with our experiments, the upper, central part of the adhered microsphere is described by a spherical cap parameterized by $(h,R,a)$ with $h$ the height of the cap, $R$ its radius, and $a$ the contact radius of the solid gel component only; 
this parameterization is consistent with the earlier TEA model \cite{Salez2013}. 
In the contact zone, around the edges of the central spherical cap, we allow for a phase-separated fluid foot of volume $\Delta V$ to be extracted from the initial gel volume $V_0 = 4 \pi R_0^3/3$. The droplet-air interface has a profile given by a shape function $\mathbf{X}=(\rho(z),z)$, specified in cylindrical coordinates, which is constrained to remain outside the spherical cap. 

To predict the configuration as a function of material parameters, we formulate and minimize the total energy $U = U_{el} + U_{st} + U_{ad}$ of the system, explicitly including bulk elastic, surface tension, and adhesion energy terms, respectively. 
The elastic energy $U_{el}$ includes both the work required to deform the initial sphere into the spherical cap and to compress the sphere to extract the fluid, respectively, 
\begin{align}
U_{el} & =\frac{1}{15}K \frac{a^5}{R_{0}^2} + \frac{B}{2V_{0}}(\Delta V)^{2},
\label{eq:elasticity}
\end{align}
\noindent where $B = \frac{1-\nu^2}{4(1-2\nu)} K$ 
is the bulk modulus of the gel elastic network. For simplicity, we use the small deformation elastic energy cost of transforming a sphere to a spherical cap \cite{hertz1881beruhrung,JKR1971,Maugis1995}, and approximate that the compression to extract fluid from the gel sphere as fully independent from the deformation to form the spherical cap, so that the overall volumetric compression does not significantly alter the stress distribution in the contact zone.

The surface tension energy $U_{st}$ encompasses the work required to increase the total surface area from that of the original sphere $A_{sph} = 4 \pi R_0^2$ to the total PDMS-air interfacial area of the adhered configuration against a surface energy per area $\gamma$. This total surface area can be obtained by integrating the cylindrically-symmetric profile $\rho(z)$ and adding the flat, bottom disk of the contact area, which has a radius $\rho(0)$. This yields a total surface tension energy contribution,  
\begin{equation}
U_{st}=\gamma \pi \left[\rho(0)^2 + 2 \int_0^h \rho(z) \sqrt{1 + \rho'(z)^2} dz - 4 R_0^2\right], 
\label{eq:surf}
\end{equation}
\noindent where $\rho'(z) = d\rho/dz$.

Meanwhile, the adhesion energy contribution accounts for the energy released by establishing an interface with adhesion energy per area $W$ over the entire combined contact area of radius $\rho(0)$: %
\begin{equation}
U_{ad}=-W\pi\rho(0)^{2}.
\label{eq:contact}
\end{equation}

To keep the model as simple as possible, we have neglected any strain dependence of gel material properties, and assumed that both the gel surface tension and adhesion energy are approximately the same as those of the free fluid phase. 
We have also neglected any surface tension between the gel and its own fluid phase, as our previous work suggests that it is very small, although nonzero \cite{jensen2015wetting}.

We define dimensionless material parameters using a scaling factor $\kappa$ with dimensions of pressure: $k=K/\kappa$, $b=B/\kappa$, $s=\gamma/\kappa R_{0}$ and $w=W/\kappa R_{0}$, of which only three are independent,
and dimensionless geometric parameters $\Tilde{h} = h/R_0$, $\Tilde{R} = R/R_0$, and $\Tilde{\rho} = \rho/R_0$.
With this substitution, 
the fully nondimensional energy functional $\Tilde{U} = U/(\kappa R_0^3)$ becomes: 
\begin{align}
\Tilde{U}= & 
\frac{k}{15} \left[  2 \tilde{R} \tilde{h} - \tilde{h}^2 \right]^{5/2} + \frac{\pi b}{24}\left[\Tilde{h}^{2}(3\Tilde{R}-\Tilde{h})-4\right]^{2} \nonumber \\
& + s \pi \left[\Tilde{\rho}(0)^{2} + 2 \int_{0}^{\tilde{h}} \Tilde{\rho} \sqrt{1 + \Tilde{\rho}'^{2}} d\tilde{z} - 4 \right] \nonumber \\
& - w \pi \Tilde{\rho}(0)^{2}
\label{eq:optimize}
\end{align}

In order to predict how individual compliant spheres will adhere, we minimize the objective functional \eqref{eq:optimize} 
with respect to $\Tilde{h}$, $\Tilde{R}$, and the shape function $\mathbf{X}$ subject to an overall volume constraint, $\int_{0}^{\Tilde{h}}\Tilde{\rho}(z)^{2}dz=\frac{4}{3}$ as well as an additional constraint, that $\left|\mathbf{X}-\mathbf{X}_{0}\right|^{2}>\Tilde{R}^{2}$ where $\mathbf{X}_{0}=(0,\Tilde{h}-\Tilde{R})$ is the center of the spherical cap. This last constraint enforces that the shape function remains spherical in the upper region of the adhered sphere, but at some point deviates away from the spherical shape to form the foot. 
Implicitly, it also ensures that the top of the foot is tangential to the spherical cap, consistent with the experimentally-observed profiles.
In a limit where the bulk modulus of the gel becomes large, the formation of a fluid foot is no longer energetically favorable. 
In this regime, our model reduces to the earlier TEA model \cite{Salez2013}, slightly modified to account for the undeformed reference state in the total surface tension energy. 
Although the XEC model only formally reduces to the TEA model in the limit of high $k$, we find that the $\Tilde{h}$ and $\Tilde{R}$ values that minimize the total energy closely coincide for both models, even though the inclusion of fluid phase separation in the XEC model modifies the energy landscape around this minimum.

We minimize \eqref{eq:optimize} using \emph{Morpho} \cite{joshi2024programmable}, a finite element program designed for shape optimization problems (see Materials and Methods; code is included in the Supporting Information). 
Predicted morphologies are displayed in Fig. \ref{fig:model}A. We use $b/k=3$, consistent with Poisson's ratio $\nu \approx 0.48$ as measured experimentally \cite{jensen2015wetting}, and choose $\kappa R_{0}=\gamma$ so that $s=1$. 
We then sweep through $0.1<k<200$ for $w=1.95$ consistent with \emph{a priori} experimental estimates. These predictions closely resemble experimentally observed shapes such as those shown above in Fig. \ref{fig:contact_geometry}A. 

Motivated by the experimentally observed dispersion of morphologies even for similar preparation conditions as shown in Fig. \ref{fig:contact_geometry}, we tested the predictions of the model by fitting each experimental droplet profile to the model using a library of simulated droplet shapes as described in Materials and Methods. For this analysis, we treat $s$ and the Poisson ratio of $0.48$ as fixed and use $k$ and $w$ as fitting parameters. For each experimental droplet, a direct search of the library, containing simulated profiles with $w\in[1.5,1.95]$ and $k\in[0.1,100]$ was performed and the closest matching droplet identified. 
We find that the fitted $w$ values are clustered around $w = 1.95$, as expected from the experiment.
Representative profile fits are shown in Fig. \ref{fig:model}B for a range of values of $k$, indicating a good level of agreement over the range of $k$ values.

We also compared the predicted $\Tilde{h}$ and $\Tilde{R}$ of the XEC model without fitting to experimental measurements. This requires that we first nondimensionalize the data with appropriate values of the surface tension and adhesion energy.
To do so, we fit the measured $h/R_0$ vs. $K$ data with the somewhat simpler TEA model to obtain average values for the surface material properties of $W_\text{fit} = 175$ mN/m and $\gamma_\text{fit} = 90$ mN/m.
We note that this surface tension is fairly high compared to that of the uncrosslinked fluid.
Historically, reported values of surface tension for soft polymer gels have varied significantly depending on the material and measurement geometry \cite{xu2016surface,jensen2015wetting,mondal2015estimation,style2013universal,jagota2012surface,Nadermann2013, park2014visualization}, which may arise as as result of strain-dependent properties via physical processes like the Shuttleworth effect \cite{Shuttleworth1950,style2017elastocapillarity,xujensen2017direct,schulman2018surface,bain2021surface,heyden2024distance}.
Additionally, other material properties may be strain-dependent, such as the elasticity, which is also not accounted for in our models.
However, since our focus here is on understanding the connection between the microscopic contact geometry, fluid phase separation, and elastocapillary physics, for simplicity we have assumed constant surface properties, and so nondimensionalize our experimental data with $k_\text{expt} = K R_0/\gamma_\text{fit}$.

In Fig. \ref{fig:model}B\textit{(inset)} we display the fitted simulation value obtained from profile matching of $k$ against the experimental value on $\log-\log$ scales, with the $k_\text{fitted} = k_\text{expt}$ line indicated. 
Overall, we see reasonable agreement over most of the measured parameter space, with significant deviations only in the highly compliant regime where the spheres are most highly deformed.  
There are several possible explanations for this deviation at high strain: 
First, the elastic energy incorporated into the XEC model is strictly appropriate only for small deformations, and could in the future be replaced with an improved large-deformation model for the compliant regime. 
Additionally, it is likely that the surface tension and/or elasticity exhibit strain-dependence, as discussed above, which would be expected to have the strongest influence in the most highly deformed spheres.

\section*{Versatile adhesion in a shallow energy landscape}

\begin{figure*}[t]
    \centering
    \includegraphics[width = 1.0\textwidth]{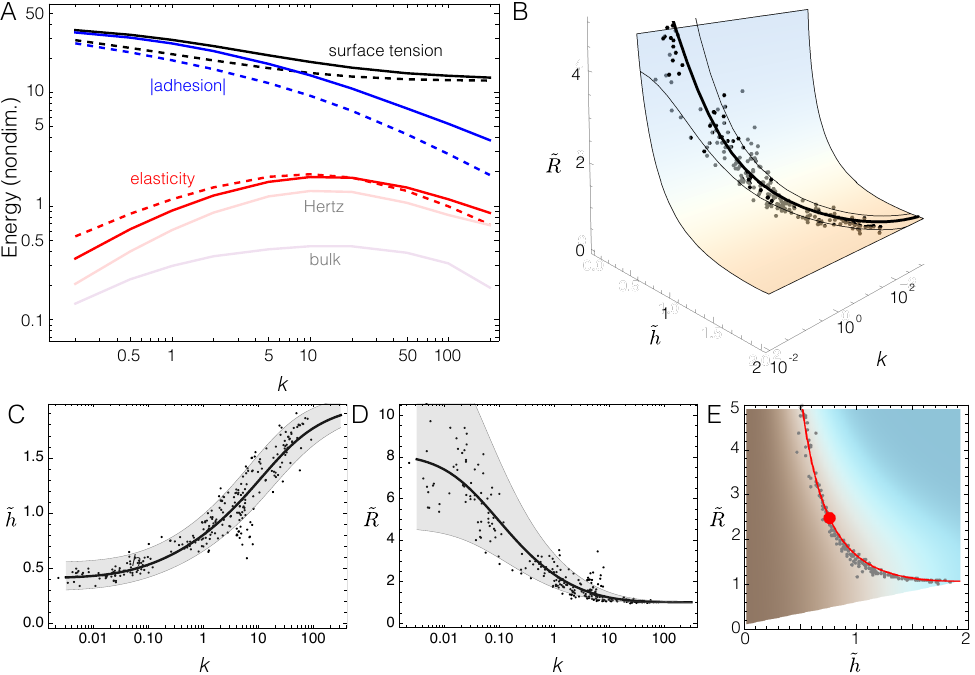}
    \caption{\textbf{Energy tradeoff and versatile energy landscape.}  
        \textit{(A)} Magnitude of energy contributions at equilibrium incorporating surface tension, adhesion and elasticity, with total, Hertz and bulk contributions shown separately. Solid lines indicate XEC model; dashed lines are TEA model.
        \textit{(B)} Energy landscape on the volume constraint manifold for TEA model as a function of $k$; thick solid lines indicate the constrained energy minimum, while thin lines indicate solutions with $10\%$ larger energy than the minimum.  
        \textit{(C,D)} Equilibrium values of $h$ and $R$ as a function of stiffness $k$ with $w=1.95$; also shown in the gray band are configurations within $1\%$ of the global minimum energy for the TEA model. \textit{(E)} Energy landscape of TEA model as a function of gel microsphere height $h$ and radius $R$ for $w=1.95$ and $k=1$. Red line indicates constant volume solutions; global minimizer is indicated with a point.}
    \label{fig:landscape}
\end{figure*}

To better understand the adhesive behavior, we now use the XEC model to examine the energetic tradeoffs made in forming the phase-separated fluid foot. 
In Fig. \ref{fig:landscape}A, we display the elastic, surface tension, and adhesive contributions to the total energy Eq. \ref{eq:optimize} of adhered spheres at equilibrium as a function of $k$ for fixed $w=1.95$, consistent with the experimental values above. 
The energy contribution magnitudes for the TEA model are also shown as dashed lines in order to highlight the differences that arise by extending the elastocapillary theory to include phase separation. 
We also display the Hertzian contact deformation and fluid-extracting bulk elastic contributions separately, showing that the Hertzian term dominates for all $k$. 

In both models, the principal tradeoff is between the surface tension and adhesion, with elasticity playing a complex role: for large $k$, elastic deformation is suppressed because it is so energetically costly, while for low $k$, elastic deformation becomes inexpensive. 
An intermediate regime emerges around $k\approx10$ where the elastic contribution to the total equilibrium energy is maximum and elasticity plays a mediating role between the surface tension and adhesion terms. 
Relative to the TEA model, the XEC model increases the adhesion and surface tension energies due to the larger surface and contact areas, while the effect on the total elastic energy is more complex due to the interaction of the bulk compression and spherical cap deformation terms.

We can also use energetic arguments to understand variations in the observed morphologies described above. While the observed morphology of each adhered sphere remains well-described by the family of shapes predicted by our extended elastocapillary model as shown in the previous section, the spheres span a wide range of equilibrium heights and adhered radii at any given $k$ value, especially in the intermediate stiffness regime.  In order to understand this apparent versatility in adhesive contact, we examine the energy landscape of solutions close to equilibrium. Doing so provides us with a sense of how small differences in experimental factors---such as potential sphere-to-sphere variations in elastic stiffness, spatial heterogeneities 
in the substrate or within the gel, hysteretic effects in establishing near-equilibrium contact line positions, or small local differences in adhesion protocol---could lead the system to adopt a state other than the global minimum for an ideal sample.

In Fig. \ref{fig:landscape}B we examine the low-energy landscape by displaying how the optimized solution on the volume constraint manifold in $(\tilde{h},\tilde{R})$ space evolves as a function of $k$  (thick black line). We also show contour lines of solutions that are $10\%$ higher in energy than the minimum (thin black lines); these indicate how the space of feasible low-energy solutions evolves as a function of $k$. We also show in Fig. \ref{fig:landscape}\textit{(C-D)} the equilibrium values of $\tilde{h}$ and $\tilde{R}$ versus $k$ separately for $w=1.95$ as solid black lines, together with the range of configurations that possess an energy within $10\%$ of the minimal value as indicated by shaded gray bands. In each plot the experimental values are shown as dots, indicating that the majority of the observed states are within the 10\% bound. 

Notably, the bounds vary strongly with $k$. For example, in the highly compliant regime as $k\ll1$, the low energy region widens. A second effect is due to the shape of the constraint manifold and how it is locally oriented around the energy minimum for different values of $k$: In the rigid regime $k\gg1$ where the solution approaches $\tilde{h}\to2$, $\tilde{R}\to1$, variations are primarily allowed in $\tilde{h}$, while for the highly compliant regime as $k\ll1$ they are mainly allowed in $\tilde{R}$. In the intermediate regime, variation in both $\tilde{h}$ and $\tilde{R}$ can occur. 

We can understand this behavior by examining the energy landscape directly. Since the TEA model is two-dimensional $(\tilde{h},\tilde{R})$, we can directly visualize it as an energy density plot, shown in Fig. \ref{fig:landscape}\textit{(E)} for the intermediate rigidity value $k=1$. 
In this plot, the optimized solution is shown as a point and is required to lie on the red line due to the TEA model's spherical cap volume constraint. 
While contour lines of energy must necessarily lie parallel to the constraint manifold by construction---this arises directly from the first-order optimality conditions---we also see that the contour lines of energy also possess similar curvature at the energetic minimum. This yields a particularly shallow potential energy landscape in the vicinity of contact equilibrium. Consequently, there exists a wide range of adhesion geometries that satisfy the volume constraint with energies only slightly above the minimum value; local perturbations to the energy landscape, such as the above-noted small differences in experimental factors, could easily trap a sphere in a nearby configuration. 

\section*{Conclusion}

Adhesion of compliant gel microspheres is fundamentally elastocapillary, smoothly transitioning from elastic-dominated mechanics for large, stiff spheres to more fluid-like, capillary-dominated mechanics with decreasing size and rigidity. Our experiments demonstrate that the gel mechanics of adhesion-induced phase separation play an essential role in establishing contact between compliant gel asperities and a non-conformal, rigid surface. We developed an extended elastocapilary (XEC) model that incorporates gel elasticity, surface tension, adhesion, and phase separation, and accurately captures not only the overall extent of deformation in adhesion like the TEA model  \cite{Salez2013} but also the specific contact morphologies observed in the experiments. The XEC model reduces to the existing TEA model when the foot vanishes, and allows us to probe the energetic tradeoffs made by the fluid foot. 

Using a library of predicted morphologies as a function of elasticity and adhesion, we were able to show that all observed adhered sphere morphologies are consistent with solutions predicted by the extended elastocapillary theory. By examining the energetic landscape around the optimal morphologies predicted by the theoretical models, we can understand the observed spread in the data, as spheres of even the same size and rigidity are able to adopt a range of heights $\tilde{h}$ and radii $\tilde{R}$ with little energetic penalty. 
This result may have practical implications for the robustness of adhesives engineered to have a rough or bumpy texture, as nearby asperities of differing size attached to the same backing material are able to deform in contact to different extents, thus cooperatively contributing to the overall contact rather than competing against each other.
Additionally, the phase-separated fluid meniscus could assist in establishing contact with particularly rough or patterned substrates with minimal additional elastic penalty. 

Our combined experimental and theoretical investigation of compliant gel microspheres across a very broad range of sizes and material properties provides unique insights into both fundamental and practical aspects of soft adhesion, and particularly emphasizes a dual role for capillarity---both solid and liquid---in rough adhesion with soft gels. 
Future experimental work will consider both the statics and dynamics of phase separation in rough adhesion over a range of gel compressibilities and varied adhesion energy between gel and substrate. The extended elastocapillary model developed here captures the essential physics of soft gel adhesion including phase separation, but still makes significant simplifying approximations by assuming linear elasticity and strain-independent material properties.  
Additional future work will particularly seek to examine elastic models valid in the large deformation limit and address the role of strain-dependent surface properties in soft adhesion, shown in earlier work to enhance the stiffness of soft contacts \cite{jensen2017strain}. 
Because soft gel asperities can deform very significantly in adhesive contact, strain-dependent surface properties could play an increasingly important role in determining their contact mechanics.

\begin{acknowledgments}
We thank Jay Thoman for useful discussions and providing the Pyrex glass cylinders, and
Hyeongjin Kim and Adam Dionne for assistance with early imaging experiments and data analysis, respectively. 
We acknowledge funding from the National Science Foundation under Grants Nos. CMMI-2129463 (KEJ) and ACI-2003820 (MQG, CJ and TJA). 
JNH received support from the Williams Summer Science Program.
EWL received support from the Williams Alumni Sponsored Internship Program. 
We also gratefully acknowledge support from a Research Corporation Cottrell Scholars Collaborative Grant.
\end{acknowledgments}

\appendix

\section{Materials and Methods}

\subsection*{Preparation of PDMS gel microspheres}
Silicone gels are prepared by mixing divinyl-terminated linear-chain polydimethyl siloxane (PDMS) (Gelest, DMS-V31) with a chemical crosslinker (Gelest, HMS-301) and platinum Karsted's catalyst (Gelest, SIP6831.2).
To aid in mixing, we predilute the catalyst at 0.05 wt\% in the polymer (i.e. by mixing 219 $\mu$L of SIP6831.2 in 400 mL of DMS-V31) and, separately, predilute the crosslinker at 10 wt\% in the polymer.
We then combine these premixes, labeled A and B, respectively, in different ratios to achieve gels with different rigidity, since the ratio of polymer to crosslinker determines the resulting gel stiffness.
After degassing the mixture in vacuum, we fabricate microspheres by emulsifying several drops of the still-fluid PDMS in 20 mL of deionized (DI) water by vortexing thoroughly several times during the work time of the gel mixture (about 10 minutes, set by the catalyst concentration).
The relatively low viscosity of the Gelest DMS-V31 (1000 cSt) relative to some commercially-available PDMS mixes facilitates the breakup into small droplets.

Because PDMS is only slightly less dense than water ($\rho = 0.97$ g/cm$^3$), and the overall volume fraction is low, the PDMS spheres easily remain dispersed in the water over time without needing any stabilizing surfactant.
We then allow the emulsified droplets to crosslink covalently into solid gel spheres at room temperature for at least 24 hours.
The resulting gel microspheres are highly polydisperse, ranging from several micrometers to around 100 $\mu$m in diameter.
A bulk sample of the same original gel mixture is reserved and cured in parallel for rigidity measurements. 

\subsection*{Measurements of gel rigidity}
We measure the rigidity of the cured gels, $K = \frac{4 E}{3 (1-\nu^2)}$ \cite{Maugis1995, Salez2013}, where $E$ is Young's modulus and $\nu \approx 0.48$ \cite{jensen2015wetting} is Poisson's ratio, by performing Hertzian flat-punch measurements on the bulk samples.
Using a TA.XTPlus Texture Analyzer (Stable Micro Systems) with a $2a = 3$ mm diameter cylindrical flat punch, we measure $F$ vs. $d$ force-indentation curves and fit the initial, linear elastic regime as Hertzian contact with $F = \frac{3}{2} a K d$.
We measure the elastic modulus of the bulk sample just prior to any experiments with its associated spheres in case of any aging of the gel over time. 
By adjusting the polymer to crosslinker ratio, we fabricate solid gels ranging from $K \approx 1$ MPa at higher crosslinker concentrations to $K \approx 0.03 $ kPa at the lowest crosslinker ratios that still form solid gels. 
We additionally performed control experiments to determine whether curing while saturated with water affects the gel stiffness, but found no significant difference in the resulting elastic moduli when identical bulk samples were cured in air or fully immersed in water.

\subsection*{Gel fluid fraction measurements}
We measure the fluid fraction in the gels by fully extracting $\sim 1$-cm$^3$ bulk samples using toluene.
In each extraction step, the samples soak in toluene in closed vials for at least two days to allow the fluid PDMS to dissolve into the toluene.
Then, we pour off the toluene---now containing dissolved liquid PDMS---into a fluid collection vial for each sample and refresh the toluene for the next soaking step.
After each extraction step, we allow the toluene to evaporate fully and weigh the total extracted fluid in each fluid collection vial.
We repeat this procedure until we have extracted all of the free fluid, determined by when any additional extracted fluid is $<$0.1\% of the original sample weight;
this usually takes 5-6 extraction steps.
Finally, we allow any remaining toluene to evaporate away from both the PDMS fluid and the solid gel network, and compare the weights of the extracted fluid and remaining solid to that of original sample to confirm that 100\% of the original mass is accounted for.

\subsection*{Preparation of rigid glass substrates}
We prepare the rigid glass substrates from bare cylindrical Pyrex glass (Corning) rod with a diameter of 3-4 mm, cleaned thoroughly with isopropyl alcohol before each experiment. 
The large-radius cylindrical geometry enables us to image adhered spheres on a clearly-visible edge that is flat parallel to the length of the rod and only slightly curved in the orthogonal direction, with a radius of curvature much larger than that of the microspheres.
We also performed extensive preliminary experiments on flat, soda-lime glass microscope slides, tilted very slightly to ensure visibility of the adhered spheres;
however, these experiments do not allow a complete mapping of the adhered sphere geometry except for the stiffest microspheres, so only the very high $K$ measurements with this adhesion geometry were included in further analyses (open symbols in Fig. \ref{fig:contact_geometry}).

\subsection*{Imaging of adhered gel microspheres}
Prior to each adhesion experiment, we vortex the cured microsphere dispersion to evenly distribute the spheres, then extract a few mL into a small puddle on a flat surface.
Because PDMS is slightly less dense than water, the microspheres slowly cream to the surface over time.
Thus, we wait several minutes before collecting spheres in order to increase our yield, but not so long that the spheres become dense enough to aggregate on the surface.
We then gently bring a rigid glass substrate (described above) into contact with the surface of the dispersion, allowing the spheres to adhere to the glass with minimal externally-applied forces.
After adhesion, we wait at least 30 minutes prior to imaging in order to allow the adhesive contacts to equilibrate and for any residual water to evaporate.
We image the adhered microspheres directly in side-view using brightfield microscopy on an inverted microscope (Nikon Ti2) using a 40x (N.A. 0.60) extra-long working distance air objective lens.

\subsection*{Image analysis of microspheres}
From raw images like those shown in Fig. \ref{fig:contact_geometry}\textit{(A)}, we map the profile of the dark edge of the flat glass and the adhered with a resolution of 100 nm using custom Matlab software developed in our earlier work to walk along lines of maximum image gradient \cite{jensen2015wetting,jensen2017strain,xujensen2017direct}.
We identify the flat glass by fitting a line to the profile far from the adhered microsphere; this becomes the $z=0$ axis by subtraction.
Note that the raw images shown in Fig. \ref{fig:contact_geometry}\textit{(A)} are cropped to focus on the sphere, so there is generally more of the surrounding flat glass surface visible in the full image.

Next, we identify the adhered microsphere profile as the region of the profile that deviates from the surrounding flat surface. 
We fit the central, spherical cap region of the microsphere profile with a circle to measure $R$, $h$, and $\Theta$, and to identify the center of axisymmetry, which becomes the $r=0$ axis in cylindrical coordinates.
We measure the total volume $V_0$ of each adhered microsphere by integrating a volume of revolution over the mapped profile from the center outwards through the end of the meniscus or ``foot.''
Because all components of these gels are non-volatile, including the PDMS free fluid phase, total volume must be conserved, and so we can calculate the original sphere radius from $V_0 = 4 \pi R_0^3/3$.
We additionally measure the local total curvature in 3D along the gel surface averaged over profile path length segments of 4 $\mu$m by fitting with a surface of constant total curvature, 
as developed in our earlier work \cite{jensen2015wetting}, again assuming axisymmetry.
(See Supporting Information Figs. S4 and S5) 
\textbf{}

\subsection*{Extended elastocapillary model optimization}

For a given set of physical parameters (k,b,w,s) we use find initial values of $\tilde{h}$ and $\tilde{R}$ determined by minimizing the energy functional \eqref{eq:optimize} assuming the absence of a fluid foot, i.e. that $\rho(z)$ remains a segment of a spherical cap; this step is equivalent to minimizing the original TEA model of Salez et al.\cite{Salez2013}. The optimized values of $\tilde{h}$ and $\tilde{R}$ are used to construct an initial finite element representation of the droplet profile, represented as a chain of linear segments with vertex positions $\mathbf{X}_{i}$. The extended elastocapillary model energy, Eq. \ref{eq:optimize}, is then minimized with respect to $\mathbf{X}_{i}$ using the following method\cite{nocedal1999numerical}: The constrained problem is first converted to an unconstrained problem using a penalty scheme where the objective function is supplemented by the constraint functions in the $L_{1}$ norm multiplied by a penalty parameter $\mu$. Optimization proceeds by minimizing the new unconstrained objective for successively increasing values of $\mu$ until an overall constraint tolerance is met. Minimization for each subproblem is performed using a limited-memory quasi-newton method\cite{nocedal1999numerical}. While other optimization schemes could be developed for this problem, the approach adopted here is fast, appears robust and allows us to efficiently explore the parameter space. All simulations are performed using the \emph{Morpho} software for shape optimization\cite{joshi2024programmable} and code to do so is provided as Supplementary Information.

\subsection*{Comparison of experimental and predicted droplet profiles} 

Droplet profiles were extracted from the experimental data set as described above and rescaled by the measured initial droplet radius $R_{0}$ prior to contact. We also prepared a library of simulated droplet profiles, already in dimensionless form, with parameters on
the interval $w\in[1.5,1.95]$ and $k\in[0.1,100]$ and with $\nu=0.48$. For every experimental and simulated droplet profile, we construct a polynomially interpolated profile $\mathbf{X}_{i}(u)$ as a function of a single parameter $u\in[0,1]$. We used the interpolated profiles to compute a similarity measure, 
\begin{eqnarray*}
S_{ij}&=&\left\langle \mathbf{X}_{i},\mathbf{X}_{j}\right\rangle \\ &=&\int_{0}^{1}\left\Vert \mathbf{X}_{i}-\mathbf{X}_{j}\right\Vert ^{2}\text{d}u,\ i\in[1,N_{expt}],\ j\in[1,N_{sim}]
\end{eqnarray*}
where $N_{expt}$ and $N_{sim}$ reflect the size of the libraries of experimental and simulated profiles respectively. We used this measure for every experimental profile to find the closest matching simulated profile using a direct search method in \texttt{Mathematica}.


%

\end{document}